\documentclass[aps,prl,twocolumn,showpacs,showkeys,amsmath,superscriptaddress,floatfix,longbibliography]{revtex4-1}

	\usepackage{amsmath}
	\usepackage{amssymb}
	\usepackage{makeidx}
	\usepackage{amsfonts}
	\usepackage[utf8]{inputenc}
	 \usepackage[T1]{fontenc}
	\usepackage[usenames,dvipsnames]{pstricks}
	\usepackage{subfigure}
	\usepackage{epsfig}
	\usepackage{pst-grad} 
	\usepackage{pst-plot} 
	\usepackage[colorlinks,hyperindex]{hyperref}
	\hypersetup
	{
		colorlinks,%
		citecolor=black,%
		linkcolor=black,%
		urlcolor=black,%
	}

\renewcommand{\vec}[1]{\mathbf{#1}}

\begin{document}

\title{Quench dynamics of a confined ultracold Fermi gas:\\Direct visibility of the Goldstone mode in the single-particle excitations}

\author{P.~Kettmann}
\affiliation{Institut f\"ur Festk\"orpertheorie, Westf\"alische Wilhelms-Universit\"at M\"unster, 48149 M\"unster, Germany}

\author{S.~Hannibal}
\affiliation{Institut f\"ur Festk\"orpertheorie, Westf\"alische Wilhelms-Universit\"at M\"unster, 48149 M\"unster, Germany}

\author{M.~D.~Croitoru}
\affiliation{Departement Fysica, Universiteit Antwerpen, 2020 Antwerpen, Belgium
}


\author{V.~M.~Axt}
\affiliation{Theoretische Physik III, Universit\"at Bayreuth, 95440
Bayreuth,
Germany}

\author{T.~Kuhn}
\affiliation{Institut f\"ur Festk\"orpertheorie, Westf\"alische Wilhelms-Universit\"at M\"unster, 48149 M\"unster, Germany}

\date{\today}

\begin{abstract}
We present a numerical study of a confined ultracold Fermi gas showing that the Goldstone mode of the BCS gap is directly visible in the dynamics of the single-particle excitations. To this end, we investigate the low-energy dynamic response of a confined Fermi gas to a rapid change of the scattering length (i.e., an interaction quench). Based on a fully microscopic time-dependent density-matrix approach within the Bogoliubov-de Gennes formalism that includes a 3D harmonic confinement we simulate and identify the emergence of the Goldstone mode in a cigar-shaped $^6$Li gas. We show that the quench leads to a low-frequency in-phase oscillation of the single-particle occupations. Complete inversion is achieved for occupations corresponding to the lowest-lying single-particle states.
\end{abstract}



\maketitle
Ultracold Fermi gases provide a unique system to test concepts of many-particle physics as well as particle theory. On the one hand, they can form both a superfluid Bardeen Cooper Schrieffer (BCS) phase and a Bose-Einstein condensate (BEC), the regimes being connected by a smooth crossover \cite{Giorgini2008Theory, Bloch2008Many}. On the other hand, the emergence of a BCS phase is associated with a spontaneously broken $U(1)$ symmetry which makes ultracold Fermi gases a convenient candidate to study the fundamental concept of spontaneous symmetry breaking (SSB) \cite{weinberg1996quantum}.

Spontaneously broken gauge symmetries and the resulting two types of fundamental collective excitations --amplitude/Higgs modes and phase/Goldstone modes-- are of fundamental interest for several fields of physics like condensed matter and particle physics. Probably the most prominent application of the concept of SSB is the Higgs mechanism \cite{Higgs1964Broken} in particle physics, where the Higgs field breaks the symmetry of the electroweak interaction and as a result the corresponding gauge bosons gain mass. In condensed matter physics SSB occurs in several systems, for example in ferromagnets (see, e.g., \cite{burgess2000goldstone}), superfluid $^3$He \cite{Paulson1973Propagation,Lawson1973Attenuation} and BCS superconductors \cite{anderson1958random, weinberg1996quantum}. In these cases, the fundamental excitations --known as magnons (i.e., spin waves), second sound (i.e., heat waves) and plasmons-- correspond to the Goldstone modes resulting from SSB.


In the field of ultracold Fermi gases the Higgs and the Goldstone mode have received great attention over the past years. The Higgs mode, i.e., the amplitude oscillation of the BCS gap, is difficult to address in experiment since it does not couple directly to external probes \cite{pekker2015amplitude}. This is why measurements of the Higgs mode have only recently been achieved for lattice superfluids \cite{bissbort2011detecting,Endres2012The/Higgs/amplitude} (in the case of BCS superconductors an experimental access has been found via THz spectroscopy \cite{Matsunaga2013Higgs,Matsunaga2014Light}), while several theoretical studies on the Higgs mode have been reported \cite{Barankov2004Collective,Barankov2006Synchronization,Yuzbashyan2006Relaxation,Dzero2007Spectroscopic,Scott2012Rapid,Bruun2014Long,hannibal2015quench}. In contrast, the Goldstone mode, i.e., the phase oscillation of the BCS gap, has been intensively studied both theoretically and experimentally (see, e.g., \cite{Kinast2004Evidence, Kinast2004Breakdown, bartenstein2004Collective, altmeyer2007dynamics,  altmeyer2007precision, riedl2008collective, baranov2000low, bruun2001low, bruun2002low, hu2004collective, heiselberg2004collective, Stringari2004Collective, grasso2005temperature, korolyuk2011density}).


However, most previous studies have been based on the collective oscillations of the trapped cloud induced, e.g., by various schemes of confinement change or by optical excitation. In contrast, we propose an alternative access to the Goldstone mode: the dynamics of the single-particle excitations of an ultracold Fermi gas induced by an interaction quench. In doing so, we show that the single-particle occupations oscillate in time with the main frequency given by the frequency of the Goldstone mode. In addition, we find a resonance of the lowest-lying single-particle excitations which leads to a full inversion of the corresponding occupations.

This full inversion opens a new and convenient experimental access to the Goldstone mode: The lowest-lying states provide the largest achievable signal amplitude in the context of the single-particle excitations and could be investigated, e.g., using RF-spectroscopy in combination with an optical control of a Feshbach resonance. On the one hand, RF-spectroscopy has already successfully been applied to probe the single-particle excitations of the ground state of an ultracold Fermi gas in the unitary regime \cite{Stewart2008Using} (an application to a dynamical situation has not been reported so far). On the other hand, the optical control of Feshbach resonances allows for an abrupt change of the interparticle interaction on a nanosecond timescale \cite{clark2015quantum}, i.e., an interaction quench. A combination of both techniques could thus provide an access to the dynamics studied in this work.

In this letter we study the dynamics of the single-particle excitations of a confined ultracold $^6$Li gas in the BCS-BEC crossover regime. In particular, we investigate the BCS side of this crossover (i.e., the Fermi wave vector $k_F$ times the scattering length $a$ is given by $0>1/k_Fa>-1$) at $T=0$. We use the approach presented in Ref. \cite{hannibal2015quench} extended by a self-consistent calculation of the chemical potential to take into account the significant change of the chemical potential during a quench in the BCS-BEC crossover regime (cf. \cite{Leggett1980Diatomic}). Like in Ref. \cite{hannibal2015quench} we calculate the dynamics of the Bogoliubov quasiparticles in the framework of the Bogoliubov-de Gennes (BdG) formalism. The investigated system is in the low-density regime, i.e., we use a short-range s-wave interaction between particles with opposite spin. The confinement is modeled by a cigar-shaped 3D harmonic potential, which in good approximation describes the standard laser confinement used in experiment \cite{Bloch2008Many}. The system is excited by an instantaneous interaction quench. The details of the formalism are described in Ref. \cite{hannibal2015quench}. Thus, here we restrict ourselves to the description of the single-particle excitations.

\begin{figure}[t]
	\centering
	\includegraphics[width=1\columnwidth]{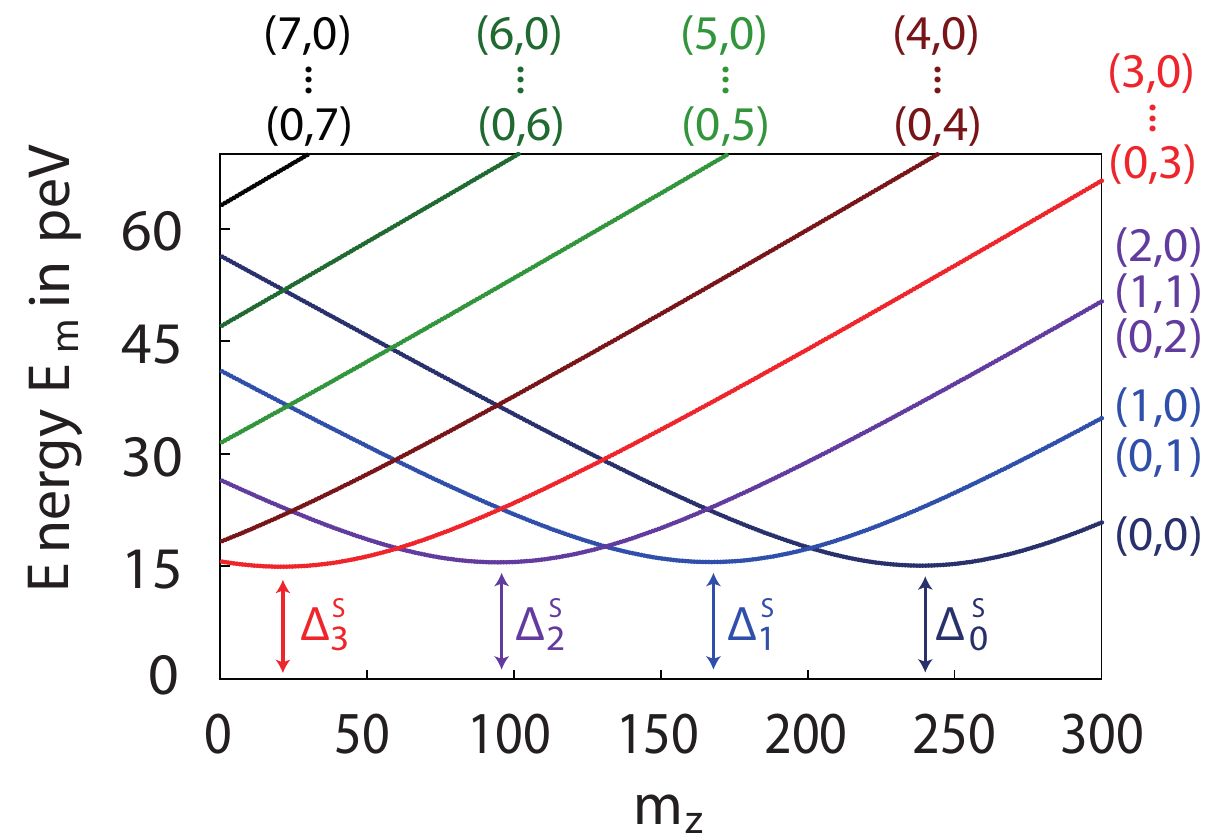}
	\caption{(color online) Single-particle energies for a strongly confined Fermi gas in a BCS phase with four atomic subbands crossing the chemical potential; $\Delta^{S}_i$ denotes the gap of the subbands with $m_x+m_y=i$ and $(m_x,m_y)$ denotes the subband index (see main text); parameters: $f_\parallel = 56\,$Hz, $f_\perp = 4\,$kHz, $N=2000$ and $1/k_Fa=-0.4$.}
	\label{fig:Eq}
\end{figure} 

The BdG Hamiltonian is diagonalized by the Bogoliubov quasiparticle transformation \citep{DeGennes1989Superconductivity}. The quasiparticles resulting from that transformation are the single-particle excitations of the BCS phase, which are created, when the system is perturbed. In the excitation picture the corresponding operators read:
\begin{align}\label{eq:B-Transformation}
	\gamma_{ma}^{\dagger} &= \int \, \left[  u_{m}(\vec{r}) \Psi_{\uparrow}^{\dagger}(\vec{r}) + v_{m}(\vec{r}) \Psi_{\downarrow}^{}(\vec{r}) \right] \, d^3 r, \nonumber \\
	\gamma_{mb}^{\dagger} &= \int \, \left[ u_{m}(\vec{r}) \Psi_{\downarrow}^{\dagger}(\vec{r}) -v_{m}(\vec{r}) \Psi_{\uparrow}^{}(\vec{r})  \right] \, d^3 r. 	
\end{align}
Here, $\Psi^{\dagger}_\sigma(\vec{r})$ is the field operator creating one atom with spin $\sigma$ at the position $\vec{r}$ and $(u_{m},v_{m})$ is the two-component wave function of a single-particle state. The index $m=(m_x,m_y,m_z)$ denotes the quantum numbers of the bare atomic states, i.e., the eigenstates of the harmonic confinement potential. The index $a$ ($b$) denotes the single-particle branch corresponding to spin-up (spin-down) particles.

The single-particle energies can be derived from the Bogoliubov-de Gennes equation \cite{Datta1999Can} and read 
\begin{equation}\label{eq:Eq}
	E_{ma}=E_{mb}=:E_m = \sqrt{\varepsilon_m^2 + (\Delta^{GS}_m)^2},
\end{equation}
where $\varepsilon_m$ is the bare atomic energy measured with respect to the chemical potential and $\Delta^{GS}_m := \langle m | \Delta(\vec{r})| m \rangle $ is the expectation value of the BCS ground-state gap with respect to the harmonic eigenstate $| m \rangle $. In the following we will analyze the structure of this energy spectrum as well as the dynamics of the corresponding excitations in detail. We will show, that after a quantum quench the occupations of the single-particle states oscillate in phase with one dominant low frequency. We will identify this frequency as the frequency of the Goldstone mode of the BCS gap. 

We investigate the dynamics of a cigar-shaped cloud. Thus, the confinement frequencies are chosen such that the gas is strongly confined in the \mbox{\textit{x-y}} plane and elongated in the $z$-direction: $f_x = f_y =: f_{\perp} \gg f_z := f_\parallel$. Therefore, the corresponding atomic energies form subbands since they are strongly separated with respect to $m_x$ and $m_y$ and comparatively dense with respect to $m_z$ (see Ref. \cite{hannibal2015quench}). According to Eq. \eqref{eq:Eq} the single-particle energies inherit this band structure, which can be seen in Fig. \ref{fig:Eq}.

\begin{figure}[t]
	\centering
	\includegraphics[width=1\columnwidth]{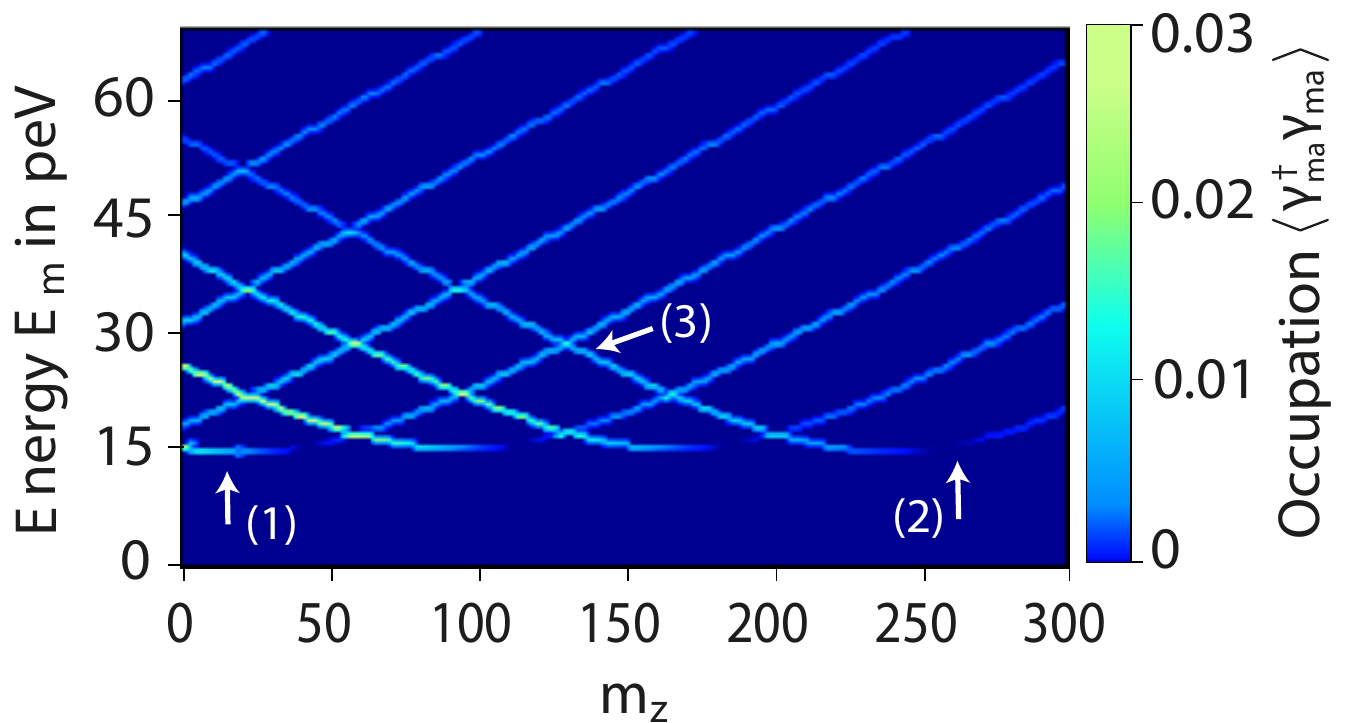}
	\caption{(color online) Single-particle occupations directly after an interaction quench from $1/k_Fa=-0.3\rightarrow -0.4$ (system parameters: see Fig. \ref{fig:Eq}); the dynamics of the marked states will be discussed below.}
	\label{fig:Sp0}
\end{figure}

There, a plot of the single-particle energies against the quantum number $m_z$ is shown for a strongly confined BCS system with the confinement frequencies given by $f_\parallel = 56\,$Hz and $f_\perp = 4\,$kHz, with $1/k_Fa=-0.4$ and with $N=2000$ atoms in the trap. One clearly observes several subbands each of which corresponds to certain sets of quantum numbers $(m_x, m_y)$, where --due to the cylindrical symmetry of the system-- each subband is $2(m_x+m_y+1)$ fold degenerate (the factor $2$ results from the degeneracy of the single-particle branches $a$ and $b$) \footnote{Actually, subbands with different and not just interchanged quantum numbers $(m_x,m_y)$ are not exactly degenerate due to slightly different subband gaps $\Delta^{S}_m$ (on the order of $0.1\,$peV for the investigated systems). For example: The subbands $(0,3)$ and $(3,0)$ are exactly degenerate, whereas the subbands $(0,3)$ and $(1,2)$ are split by $\sim 0.5 \,$peV. However, in the presented calculations and with the assumed experimental accuracy this splitting is not resolved.}. Furthermore, the four subbands with the lowest sets of quantum numbers $(m_x,m_y)$ show minima when the corresponding atomic subbands cross the chemical potential, i.e., at different values for $m_z$. The states located at the minima thus lie in close vicinity to the chemical potential and contribute strongly to the BCS pairing (the expectation values $\Delta^{GS}_m$ corresponding to these states, i.e., the subband gaps, will be denoted as $\Delta^{S}_i$ with $i=m_x + m_y$ the subband index; see Fig. \ref{fig:Eq}). The higher atomic subbands with $m_x + m_y \geq 4$ do not cross the chemical potential. The corresponding single-particle subbands therefore do not exhibit any minima.

Since the single-particle operators of Eq. \eqref{eq:B-Transformation} are defined in the excitation picture all energy states of Fig. \ref{fig:Eq} are not occupied in the ground state before the quench. However, immediately after the quench all single-particle states are occupied with the corresponding single-particle expectation values given by \cite{hannibal2015quench}:
\begin{equation} \label{eq:excitation}
\Big<\gamma_{ma}^{\dagger}\gamma_{ma}^{}\Big>\big|_{t=0} = \Big<\gamma_{mb}^{\dagger}\gamma_{mb}^{}\Big>\big|_{t=0}	= \left(v_m \tilde{u}_m - u_m  \tilde{v}_m\right)^2,
\end{equation}
where $(u_m,v_m)$ denotes the wave functions of the ground state after the quench and $(\tilde{u}_m,\tilde{v}_m)$ the ones before the quench.
The corresponding excitation spectrum is shown in Fig. \ref{fig:Sp0}. There, an energy- and $m_z$-resolved plot of the occupation of the single-particle states is shown for the same system as in Fig. \ref{fig:Eq} directly after a quench from $1/k_Fa= -0.3$ to $1/k_Fa= -0.4$.  One clearly observes that the low-lying single-particle states are occupied by the quench where the strength of the occupation decreases with increasing energy. Only excitations slightly above the subband minima are suppressed due to the symmetry of the single-particle wave functions. Furthermore, one can see that the left wings of the single-particle subbands are more strongly occupied than the right wings. This asymmetry is a result of the change of the chemical potential  \footnote{When --in the calculations-- the chemical potential is kept fixed during the quench the states directly at the minima are not occupied. I.e., the shift of the occupation minimum to higher energies is due to the change of the chemical potential.}. At fixed energy $E_m$ the states with small quantum numbers $m_{x/y}$ --i.e., those corresponding to lower atomic subbands and thus with higher values of $m_z$-- are less occupied than those with large $m_{x/y}$. However, the latter effect is due to the degeneracy of the subbands which is larger for states with higher quantum numbers $m_{x/y}$.

\begin{figure}[t]
		\includegraphics[width = 1 \columnwidth]{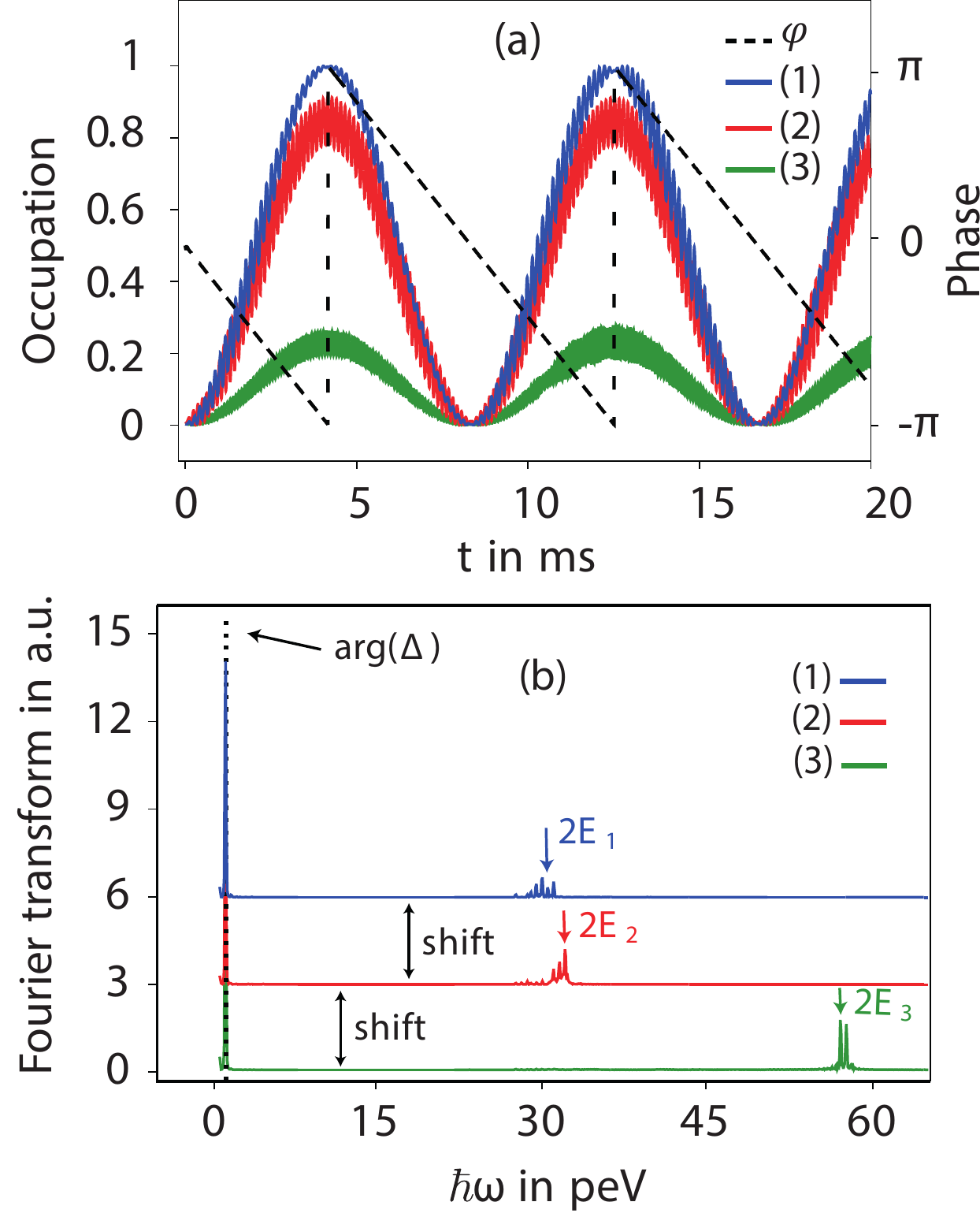}
		\caption{(color online) (a) Dynamics of three particular single-particle occupations, one state of higher energy (3) and two from subband minima (1), (2) (see Fig. \ref{fig:Sp0}). (b) Fourier transform of the functions in (a).}
		\label{fig:gamma_t}
\end{figure}

The excitations created by the quench are of the order of $10^{-2}$ and thus rather weak. Nevertheless, during the temporal evolution after the quench much larger occupations of the order of 1 are created. In the following we will show this explicitly for three particular single-particle states [marked as (1), (2) and (3) in Fig. \ref{fig:Sp0}], one from the minimum of the subband (1,2) ($E_m = 15.0\,$peV; blue line), one close to the minimum of the subband (0,0) ($E_m = 15.9\,$peV; red line) and one at a higher energy in the subband (0,0) ($E_m = 28.6\,$peV; green line). Furthermore, we will compare these dynamics to the dynamics of the complex phase of the BCS gap $\varphi := \mathrm{arg}\big(\Delta(\vec{r})\big)$ (dashed line), which turns out to be independent of the position $\vec{r}$.

The dynamics of these three single-particle occupations is shown in Fig. \ref{fig:gamma_t} (a) for the first 20$\,$ms after the quench. We clearly observe that all occupations oscillate in phase with one dominant low frequency. The states (1) and (2) have a large amplitude of the order of 1 while the amplitude of state (3) is much smaller. Furthermore, the three occupations each exhibit an individual weak higher-frequency component which has the largest frequency for the state (3) of high energy. However, we find that the amplitude of the higher-frequency component increases with decreasing the scattering length, i.e., when approaching the BCS regime with $k_F a < -1$.

A comparison of the single-particle dynamics with the dynamics of the phase of the gap [Fig. \ref{fig:gamma_t} (a); dashed line] shows that the dominant low oscillation frequency originates from the Goldstone mode of the gap: The phase of the gap decreases linearly in time with a rate corresponding to the same low frequency as the single-particle occupations. Our calculations thus show, that the Goldstone mode of the BCS gap is reflected in the excitation dynamics of a confined ultracold BCS Fermi gas after a quantum quench.

Figure \ref{fig:gamma_t} (b) shows the Fourier spectrum of the dynamics of Fig. \ref{fig:gamma_t} (a). Again, we observe that the dominant low frequencies of the occupations and the phase of the BCS gap exactly match. However, the origin of the higher frequencies in the excitation dynamics can now be seen as well: Besides the dominant low-frequency components each spectrum exhibits a series of weak peaks at approximately twice the energy of the corresponding single-particle state. Thus, the higher-frequency components result from an eigenoscillation of the single-particle occupations. The oscillation of the single-particle excitations is thus composed of two main frequency components, one originating from the Goldstone mode of the BCS gap and the other being the eigenoscillation of the single-particle occupation. In Ref. \cite{hannibal2015quench} a sum of all eigenoscillations was shown to result in the Higgs mode of the gap. The higher-frequency components can thus be understood as fragments of the Higgs mode of the BCS gap.

\begin{figure}[t]
		\includegraphics[width = 1\columnwidth]{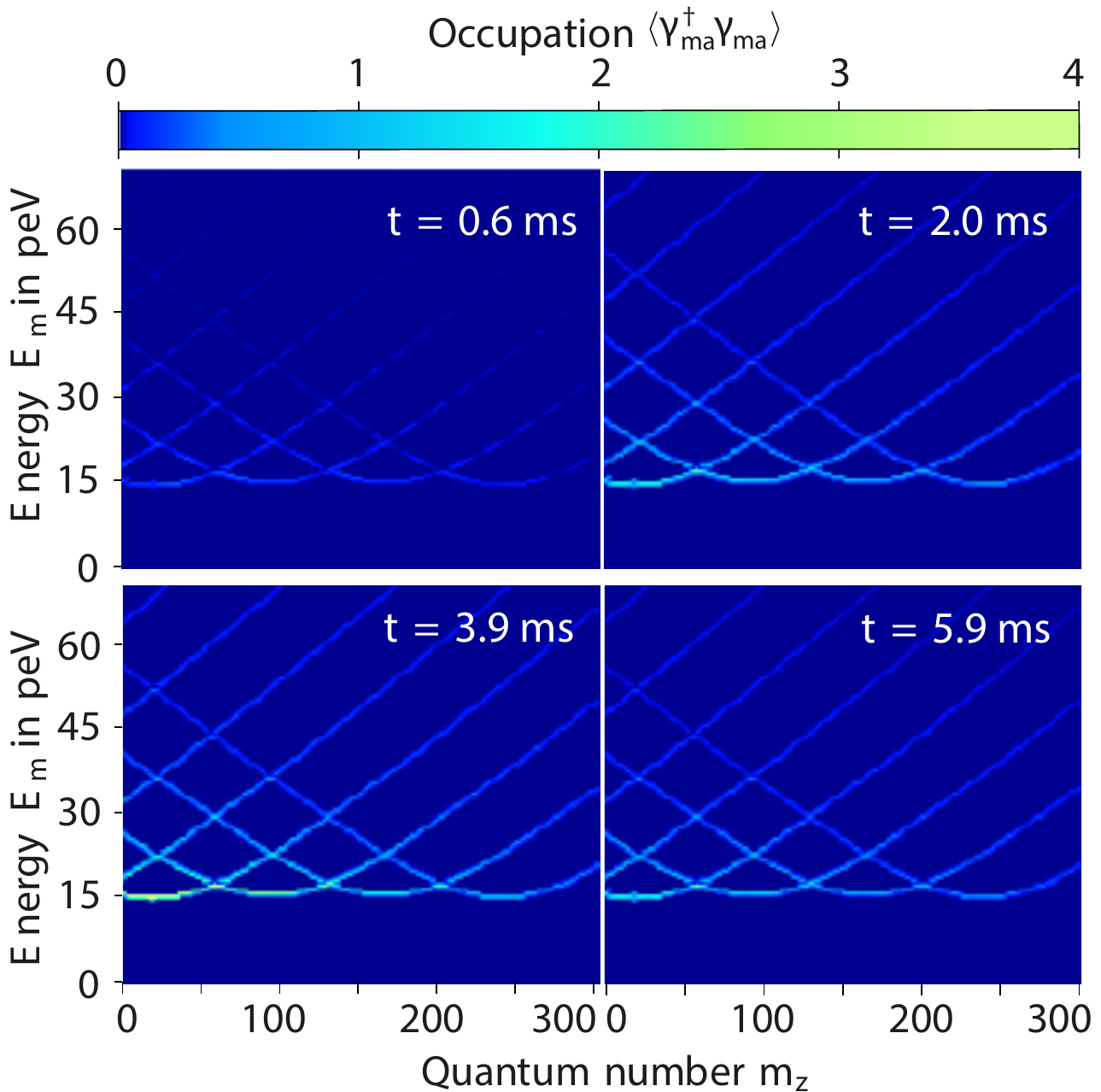}
		\caption{(color online) Single-particle occupations at different times after the interaction quench; parameters: See Fig. \ref{fig:Eq}. }
		\label{fig:spec_t}
\end{figure}


In Fig. \ref{fig:spec_t} the impact of the Goldstone mode on the whole single-particle spectrum is shown in terms of snapshots of the occupation spectrum for a series of time steps after the quench. Again, the single-particle occupations are plotted against the quantum number $m_z$ and the excitation energy, like they could be measured by angle- and momentum-resolved RF spectroscopy \cite{Stewart2008Using}. The first snapshot corresponds to the time $t=0.6\,$ms and thus directly follows the situation of Fig. \ref{fig:Sp0} but with a different scaling of the color axis. Therefore, at $t=0.6\,$ms the higher-energy excitations can hardly be seen. However, going on in time we observe that all occupations increase in phase until the time $t=3.9\,$ms, where the maximum occupation of all states is reached. Afterwards the occupations decrease until the initial situation is reached again. Thus, Fig. \ref{fig:spec_t} is an illustration of the in-phase oscillation of all single-particle occupations due to the Goldstone mode. 

In addition, --neglecting the contributions from the eigenoscillation-- the snapshot for $t=3.9\,$ms provides a map of the amplitude of the single-particle oscillations since here the dominant low-frequency part of all occupations exhibits its maximum value. On the basis of this amplitude map one observes, that the amplitude distribution reminds of the situation in Fig. \ref{fig:Sp0}: The amplitude is large for states with low quantum number $m_z$ and low energy $E_m$ and decreases with increasing values of $m_z$ and $E_m$. While the occupations of the states slightly above the minima of the excitation energies are suppressed with respect to the initial excitation, they nevertheless exhibit large oscillation amplitudes. In fact, directly at the minimum of the subbands with $m_x+m_y=3$ the oscillation amplitude is 4, decreasing by 1 for every next lower subband $m_x+m_y$. This dependence is again due to the $(m_x+m_y+1)$-fold degeneracy of the subbands (only the particle-like excitations of the single-particle branch $a$ are shown). I.e., the oscillation amplitude of each individual single-particle occupation at a subband minimum is 1. Thus, the single-particle occupations at the subband minima exhibit a full inversion.


In conclusion, we have calculated the nonequilibrium dynamics of the single-particle excitations of a confined superfluid $^6$Li gas after an interaction quench. We have shown that the single-particle occupations oscillate in time with one dominant low frequency. By comparing this frequency component to the phase oscillation of the BCS gap we have traced it back to the Goldstone mode of the BCS gap. We have shown that the lowest-lying single-particle states exhibit a full inversion due to this mode. Furthermore, we have observed a weak higher-frequency contribution in the dynamics of the single-particle occupations. We have identified its frequency as twice the energy of the corresponding single-particle state and thus, we have connected it to the Higgs mode of the BCS gap.

\end{document}